# Radical-Induced Changes in Transition Metal Interfacial Magnetic Properties: A Blatter Derivative on Polycrystalline Cobalt

Ewa Malgorzata Nowik-Boltyk,[a] Tobias Junghoefer,[a] Erika Giangrisostomi,[b] Ruslan Ovsyannikov,[b] Chan Shu,[c] Andrzej Rajca,[c] Andrea Droghetti,[d] M. Benedetta Casu*[a]

---

[a] E. M. Nowik-Boltyk, T. Junghöfer, M. B. Casu
Institute of Physical and Theoretical Chemistry
University of Tübingen
72076 Tübingen, Germany
E-mail: benedetta.casu@uni-tebingen.de

[b] E. Giangrisostomi, R. Ovsyannikov
Institute Methods and Instrumentation for Synchrotron Radiation Research
Helmholtz-Zentrum Berlin,
12489 Berlin, Germany

[c] C. Shu, A Rajca
Department of Chemistry
University of Nebraska
Lincoln, NE 68588, United States

[d] A. Droghetti
School of Physics and CRANN,
Trinity College, the University of Dublin,
D02, Ireland

Abstract

In this work, we study the cobalt/radical interface obtained by depositing a monolayer of a derivative of the Blatter radical on polycrystalline cobalt. By examining the occupied and unoccupied states at the interface, using soft X-ray techniques, and combining them with ab initio calculations, we can completely determine the electronic structure at the interface, simultaneously, on both the molecular and ferromagnetic sides, and thus obtain a full understanding of the magnetic properties at the interface. We find that the molecule is strongly hybridized with the surface. Changes in the core level spectra reflect the modification of the molecule's and cobalt's electronic structure induced by the bonding across the interface. This causes a decrease in the magnetic moments of the cobalt atoms bonded to the molecules that lose their radical character. Our method allows screening, beforehand, the organic/ferromagnetic interfaces in view of potential applications in spintronic.

Introduction

Organic/inorganic interfaces have been investigated for at least three decades [1] as they are at the core of the technological developments driven by the miniaturization of materials and devices towards the nanoscale limit. In particular, in the field of spintronics, the study of organic/ferromagnetic metal interfaces initially acquired importance in order to understand spin tunneling and spin injection from ferromagnetic electrodes into molecules.[2] Yet, the interest in these interfaces now encompasses many aspects of magnetism, with a variety of effects that have been reported to emerge when an organic layer is deposited on an ultra-thin ferromagnetic film.[2c, 2g, 3]

The organic layer at an organic/ferromagnetic metal interface often acquires a spin polarization because of the chemical bond with the ferromagnetic film.[4] In turn, the electronic and magnetic properties of the ferromagnetic film are also significantly modified.[5] A paradigmatic example is the fullerene ($C_{60}$)/cobalt interface. On one side of such interface, the $C_{60}$ layer can be used to spin filter electrons,[6] while, on the other side, the Co magnetic moments are reduced,[6] and the Co film's magnetic anisotropy and hardness are enhanced compared to the clean Co film.[7] Similar or even larger modifications of the magnetic properties of Co films have also been reported with tris(8-hydroxyquinoline) metal ($Mq_3$) molecules.[8]

The goal of our work is to go beyond the most studied molecules, $C_{60}$ and $Mq_3$, by considering yet-unexplored interfaces formed between organic radicals and Co. To achieve this goal, we propose an original and powerful approach to characterize at the same time the electronic properties of both the molecular and the metallic side of interfaces, whereas most studies to date have focused on either of the two separately. Thin films of purely organic radicals (i.e., organic molecules that carry an unpaired electron and, thus, a magnetic moment[9]) have attracted attention for many applications.[10] Recently, it was demonstrated that they can be evaporated under controlled conditions in ultra-high vacuum,[11] thus enabling the use of characterization techniques complementary to the traditional ones in the field of radical chemistry. For instance, soft X-ray spectroscopy combined with concepts originally stemming from the surface science of closed-shell systems, make it possible to access radical/inorganic interfaces and the therein occurring phenomena.[9c, 11-12]

In this work, we study the cobalt/radical interface obtained by depositing a derivative of the Blatter radical[13] (Blatter-pyr, Figure 1) on polycrystalline cobalt. In the polycrystalline form, cobalt maintains its magnetic properties[14] broadly used in spintronic devices. Blatter-pyr, obtained by fusing a Blatter radical to a pyrene unit, is a chemically and thermodynamically stable purely organic radical[13] that can be evaporated in ultra-high vacuum (UHV), under controlled conditions.[12c]

We investigated the occupied and unoccupied states at the interface, by taking advantage of the element-sensitivity of the chosen soft X-ray techniques and combining them with *ab initio* calculations. The results indicate that the molecule is strongly hybridized with the cobalt surface. Specifically, we find that the core level spectra reflect a reduction of the magnetic moments of the cobalt atoms, bonded to the molecules, which, in turn, lose their radical character.

Our work presents an original point of view for investigating not just the radical/cobalt interface, but also many other interfacial systems, because the measurements give access to both the molecular and the metallic side at the same time, with the advantage of minimizing the risk of potential discrepancies in the results which can instead occur when different techniques, working under different conditions, are used for the two subsystems.

**Results and discussion**

We evaporated Blatter-pyr on polycrystalline cobalt using organic molecular beam deposition (OMBD) which allows precise control of the evaporation parameters.[15] In this experiment, this also minimizes the possible contamination of the cobalt substrate and avoids generating a non-defined interface.

The films were investigated by X-ray photoelectron spectroscopy (XPS) which is a powerful element-sensitive analytical method.[16] Its signal is proportional to the element concentration and their chemical state in the investigated systems and coupled with *ab initio* calculations gives a clear insight into their electronic structure, also at the interface. Introducing this technique to the investigation of radical thin films revealed its ability to identify whether the unpaired electron of the radical is involved in charge transfer, with or without chemisorption, losing its imparity and, consequently, with the molecule losing its radical character.[11, 17]

Looking in detail at the C 1s and N 1s core-level spectra, we found that the main lines have a complex shape that depends on the different chemical environments of each element in the radical (Figures 1a and 1b). XPS C 1s core level spectra of the thick films of Blatter-pyr deposited on polycrystalline cobalt (Figure 1a) are characterized by a main line at around 284.6 eV which is attributed to photoelectrons emitted from the carbon atoms of the aromatic sites (C-C and C-H bound carbons), and a second feature at higher binding energies (285.8 eV) due to contributions of photoelectrons from the carbon atoms bound to nitrogen atoms (C-N).[12c] Accordingly, the N 1s core level spectra (Figure 1b) are characterized by three contributions (at around 398.3, 399.2, and 401.0 eV with $\Delta E_{N1-Nrad} = 2.7$ eV, $\Delta E_{N1-N2} = 1.8$ eV, and $\Delta E_{N2-Nrad} = 0.9 eV$) that correspond to the three different nitrogen atom chemical environments, as expected for an intact Blatter-pyr.[12c] In addition to the effects due to the different chemical environments, we also observe the presence of several satellite features (in this work indicated with $S_i$, i=1,2,3…), the so-called shake-up satellites, which are typical in photoemission and appear as an effect of the relaxation processes, caused by the core hole left behind by the photoemitted electron.[18] Their intensity has to be taken into account when calculating the stoichiometry of the investigated systems by using XPS (see the Supporting Information).[16, 19]

The stoichiometry of the films is further proved by using a well-established fit routine,[12c, 20] systematically correlated with electron paramagnetic resonance (EPR) results on a variety of different radicals.[11-12, 20a, 21] The fit indicates that all components are stoichiometrically meaningful (Tables S1-S2 in the Supporting Information). Thus, we can conclude that the XPS spectra are concomitant with those of thin films with the expected EPR pattern corresponding to intact Blatter-pyr.[12c, 20a, 21c]

The thick film spectra are not influenced by interface phenomena. Thus, to understand in detail the interface we have performed a thickness dependent investigation of the core levels spectra. Comparing the spectra of the thicker films with those of the interfacial layer (i.e., the first organic layer on cobalt formed by the molecule directly in contact with the metallic ferromagnet) and looking at the thickness-dependent behavior of the spectroscopic lines, we observe pronounced changes. In particular, the experimental spectroscopic lines of the interfacial radical layer are broadened and shifted towards lower binding energies (~ 1 eV). The fit analysis applied to the thickness-dependent spectra shows how the intensities and the binding energies of the single contributions evolve versus thickness (Figure 2). Also, the intensity of the satellite features is higher. These phenomena indicate a strong interaction, of a chemical nature, between the Blatter-pyr and the cobalt substrate. The radical character in films of intact molecules is represented by the feature at lower binding energy (Nrad in Figure 1) and correlates with its intensity.[12c, 21c, 22] This feature has lower intensity in the interfacial film N 1s core level spectrum than in the thicker films (Figure 1d, compare with Figure 1b) revealing only a residual radical character. This is because the chemical environment of molecules in contact and bonded with the cobalt surface is significantly different than in the volume of the radical films (Figures 1c and 1d, the interfacial features are indicated with an asterisk to underline their different chemical environment and Table S4 in the Supporting Information).

The finding that Blatter-pyr is chemisorbed on cobalt is further supported by the Near Edge X-ray Absorption Fine Structure (NEXAFS) spectra measured on the interfacial layers (Figure 3) at the C K edge. The signal arises from transitions from the C 1s core levels to the unoccupied states. The spectra are substantially different from the spectra of the thick films (Figure S1),[9c, 22] lacking the pre-edge feature attributed to the transitions from the C 1s core levels to the singly unoccupied molecular orbital (SUMO), and the peaks in the 288-290 eV photon energy region.[9c, 22] This result suggests that the SUMO becomes occupied due to the occurrence of charge transfer from the cobalt substrate to the radical film. It corroborates the XPS results, evidencing the chemisorption of the radical on the cobalt substrate. We also observe that the NEXAFS spectra for two different polarizations of the incident light with respect to the substrate do not show any strong dichroism,[22] revealing a high degree of azimuthal disorder in the interfacial layer.

Combining XPS and NEXAFS has provided a detailed description of the phenomena ongoing at the interface influencing the radical character of the first molecular layer deposited on cobalt. Now, we can take advantage of the element-sensitivity of X-ray techniques simultaneously by looking at the interface from the molecular and the cobalt side: we measured the NEXAFS Co $L_{2,3}$ edge spectra (corresponding to transitions from the Co 2p core level states to the 3d unoccupied states) before and after the deposition of the radical films (Figure 4).

In literature spectra are compared to understand the influence of the alloying, oxidation, or ligands on the cobalt bulk.[23] Conversely, here we use this approach to investigate a hybrid organic radical/inorganic interface.

Cobalt is a highly reactive material, which leads to the formation of various oxides and hydroxides if left exposed to ambient conditions or oxygen contaminants in high vacuum.[23g, 24] Therefore, it is important to rule out oxidation as a cause for the potential changes. The cobalt $L_3$ absorption edge shows distinctive features, depending on the oxidation state and the ligand field splitting (if applicable), such as peak splitting and shoulders in the peaks.[23f, 23g, 25] The line shape of the cobalt NEXAFS spectra after deposition does not deviate prominently from that of the clean cobalt surface (Figure 4). Neither splitting nor additional shoulders are present in the absorption spectra of the Blatter-pyr/cobalt interface after film deposition. This is a clear indication that no significant oxidation of the cobalt occurs upon radical deposition.

The signal difference evidences a slight increase in intensity after film deposition (red curve, Figure 4). The enhanced intensity indicates a lower electronic population in the final states after deposition,[3c, 23f] concomitant with cobalt-to-radical charge transfer as evinced with XPS and NEXAFS at the molecular side.

Taking inspiration from the data analysis performed on electron energy loss spectroscopy (EELS),[23a-d] we have further developed and applied it to the NEXAFS spectra to quantify the differences in the spectra before and after evaporation. We have calculated the $I(L_3)/I(L_2)$ ratio, the branching ratio $I(L_3)/I(L_2)+I(L_3)$, and the normalized white lines (i.e., $I(L_2)+I(L_3)$ normalized to the continuum),[23c, 23d] (being $I(L_3)$ and $I(L_2)$ the peak areas of the Co $L_2$ and Co $L_3$ edges, see Table S5 and the Supporting Information for definitions and details). We find that the $I(L_3)/I(L_2)$ ratio is lower after the layer deposition, indicating lower d orbital occupancy.[23a, 23c, 23d, 23f, 23g] Considering methods previously used for EELS spectra;[23c, 23d] it is also possible to obtain a numerical estimate of the d orbital occupancy. We find that it is lower after chemisorption of the Blatter-pyr (see Table S5). The analysis of the spectra reveals a consistent pattern: the cobalt surface loses charge to the radical, as indicated by all the calculated values (Table S5).

At this point, the question is how the interaction with the radical monolayer and the consequent charge transfer influence the magnetic properties of the cobalt surface. We answer this question through a detailed examination of the electronic configuration of the unoccupied states in the cobalt substrate, both pre- and post-evaporation of the interfacial film: we have calculated the difference in the magnetic moment before and after deposition utilizing calculated the $I(L_3)/I(L_2)$ ratio, the branching ratio $I(L_3)/I(L_2)+I(L_3)$, and the normalized white lines.[23a, 23b, 26] We could determine a difference in magnetic moment of -0.05 $\mu_B$ after evaporation (see the Section Methods in the Supporting Information). Although this value is a simple estimation what is important is the tendency identified by the calculation, i.e., the cobalt magnetic moment is lower after evaporation.

In pursuit of further exploring this aspect, we returned to XPS (because of its higher surface sensitivity in comparison with NEXAFS) and directed our focus towards analyzing the cobalt 3s core level spectra, characterized by the multiplet splitting that may occur when unpaired electrons occupy valence states.[27] We fitted the Co 3s core level spectra before and after the deposition of the interfacial film, by using a Doniach-Sunjic line shape (Figure 5).[27] We find that the doublet binding energy separation is around 3.5 eV, in agreement with the literature for polycrystalline cobalt and cobalt compounds.[28] The energy separation between the two peaks correlates with the 3s-3d exchange interaction energy, while the peak intensities of the doublet features correlate with the total spin and, thus, with the magnetic moment.[27, 29] We calculated the magnetic moment, obtaining 1.79 $\mu_B$ and 1.50 $\mu_B$ before and after evaporation (see the Supporting Information). Thus, by examining both the occupied and unoccupied states, we come to the result that the cobalt magnetic moment decreases due to the chemisorption of the Blatter-pyr. Here, we like to mention that this method gave good values of the magnetic moment in cobalt-based materials, chromium compounds, and iron alloys,[27-30] however some critical issues have been reported in other cases.[31] In part, the discrepancy of the published results is given by the difficulty of fitting the 3s line. Differences in the fitting procedure might lead to different values of the magnetic moment. In our opinion, these values must be considered as tendencies, useful to make comparisons among different chemical conditions of the cobalt surface, rather than to be taken as absolute values.

To support and further elucidate the experimental results we performed *ab initio* calculations based on Density Functional Theory (DFT) following the same approach used in other studies about radical molecules and their adsorption on surfaces.[12b, 21c, 32] Blatter-pyr is a planar molecule, except for the phenyl ring bonded to N1, which forms a dihedral angle, $\vartheta = 44.3^o$, with the Blatter core because of the steric interaction with the pyrene unit. The electronic structure is typical of a radical with spin S=1/2.[9c, 12b, 21c, 22, 32a, 33] It presents an unpaired electron and, therefore, a singly occupied (and a singly unoccupied) molecular orbital SOMO (and SUMO) (Figure S2 in in the Supporting Information). These are mostly localized on the three nitrogen atoms (N1, N2, N3=Nrad) of the Blatter core, with a small contribution from the fused pyrene unit, as seen in the spin density isosurface in Figure 6a. The results are overall consistent with previous DFT calculations for the same molecule.[9c, 21c, 22, 33c] The simulated N 1s core level spectrum is displayed in Figure 6a. It resembles the experimental one for thick layers. We observe three main peaks that correspond to the three different N atoms. The peak at the lowest binding energy is associated with N3 (=Nrad), the one at the highest binding energy to N1, and the one at the intermediate binding energy to N2. All peak binding energies are underestimated by about 20 eV compared to the experimental values because of the employed approximations (see computational methods in the Supporting Information). Despite that, we observe that the energy splitting of the three peaks, $\Delta E_{N1-N3} = 2.9$ eV, $\Delta E_{N1-N2} = 1.9$ eV, and $\Delta E_{N2-N3} = 1.0$ eV, agrees very well with the experimental results. Thus, the calculations confirm the correct interpretation of the core spectra provided above.

The cobalt/molecule interface is modeled by placing one molecule on a Co slab. We consider the Co (0001) surface, which is the prevalent surface in our polycrystalline samples.[34] Here we only consider one adsorption geometry, shown in Figure 6d. Different geometries are discussed in the Supporting Information. After the geometry optimization, the calculated average adsorption distance is about 2.1 Å which is typical for organic molecules strongly chemisorbed on ferromagnetic transition metals.[4a, 8c] The radical nature of the molecule is not preserved. The calculated magnetic moment of the Blatter core is only about $0.12\ \mu_B$. The molecular features, such as the SOMO and the SUMO, in the electronic structure near the Fermi level disappear as a result of the hybridization with Co, and they merge into to very broad states (see Figure S3 in the Supporting Information), similar to those found for other organic molecules on ferromagnetic transition metals.[4a] In turn, the magnetic moments of Co atoms bonded to the molecule are reduced from about $1.7\ \mu_B$, the value for the clean surface, to $0.25\ \mu_B$, reproducing the tendency of the experimental observation.

The simulated N 1s core level spectrum of the molecule on the Co slab is shown in Figure 6c. The energy splitting between the three characteristic peaks is strongly reduced compared to that seen for the isolated molecule. We obtain that $\Delta E_{N1-N3} = 2.0$ eV, $\Delta E_{N1-N2} = 1.4$ eV, and $\Delta E_{N2-N3} = 0.6$ eV. This explains the broadened line shapes of the N 1s core level curve that we observed experimentally at the interface (Figure 1d). Overall, our DFT results fully support the arguments used to interpret the experiments. The molecule is strongly hybridized with the surface, and the change in the core level spectra reflects the modification of the molecule's and cobalt's electronic structure induced by the bonding across the interface.

**Conclusions**

We evaporated Blatter-pyr on polycrystalline cobalt. By working in UHV, we ensure a controlled environment that minimizes the impact of contaminants on our results. The deposition of Blatter-pyr on cobalt influences the cobalt electronic structure and the magnetic moment through the hybridization between the molecular and the atomic orbitals.

We find that Blatter-pyr loses its radical character. In this study, this is irrelevant because our focus is on the interface as a system, and specifically on investigating the possibility of influencing cobalt properties at the interface by depositing the radical.

Future work might envisage the necessity to have an intact radical decorating the cobalt surface. This can be achieved by appropriately designing the radical molecule with functional groups that attach to the ferromagnetic surface hindering the chemisorption of the radical moiety, as already shown, for example, on gold surfaces.[20b, 32a, 35]

We show that it is possible to predict the impact of the organic interfacial layer on the magnetic properties of the polycrystalline cobalt substrate, analyzing in extreme detail X-ray photoemission and absorption data, supported by *ab initio* calculations. Despite their importance for spintronics, the number of investigated organic materials on cobalt remains quite limited because describing the electronic and magnetic properties of such an interface requires a large variety of different techniques and it is very time consuming. In this respect the versatility of our approach plays an important role, allowing preselecting, beforehand, organic/ferromagnetic and radical/ferromagnetic interfaces that have suitable properties, identifying the most promising for spintronic applications.

**Supporting Information**

The authors have cited additional references within the Supporting Information.[20b, 23c, 23d, 36]

**Acknowledgments**

The authors would like to thank Helmholtz-Zentrum Berlin (HZB) for providing beamtime at BESSY II, and Dr M. Pink for the X-ray structures of the radicals. Financial support from HZB, the German Research Foundation (DFG, contract CA852/11-3, project number: 394233453), the Federal Ministry of Education and Research (BMBF) in the framework of the program "Erforschung von Universum und Materie – ErUM-Pro" (RadicalQuantum, Grant number 05K22VT1), Science Foundation Ireland, the Royal Society through the University Research Fellowships URF-R1-191769, and the Chemistry Division of National Science Foundation (grants CHE-1955349 and CHE-2247170) is gratefully acknowledged. This work was supported by the European Union's Horizon 2020 Research and Innovation program under grant agreement no. 965046, FET-Open project Interfast (Gated INTERfaces for FAST information processes).

**Keywords:** radical thin films • cobalt • interface • density functional theory • soft x-rays

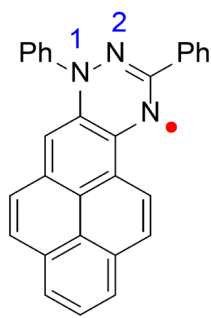
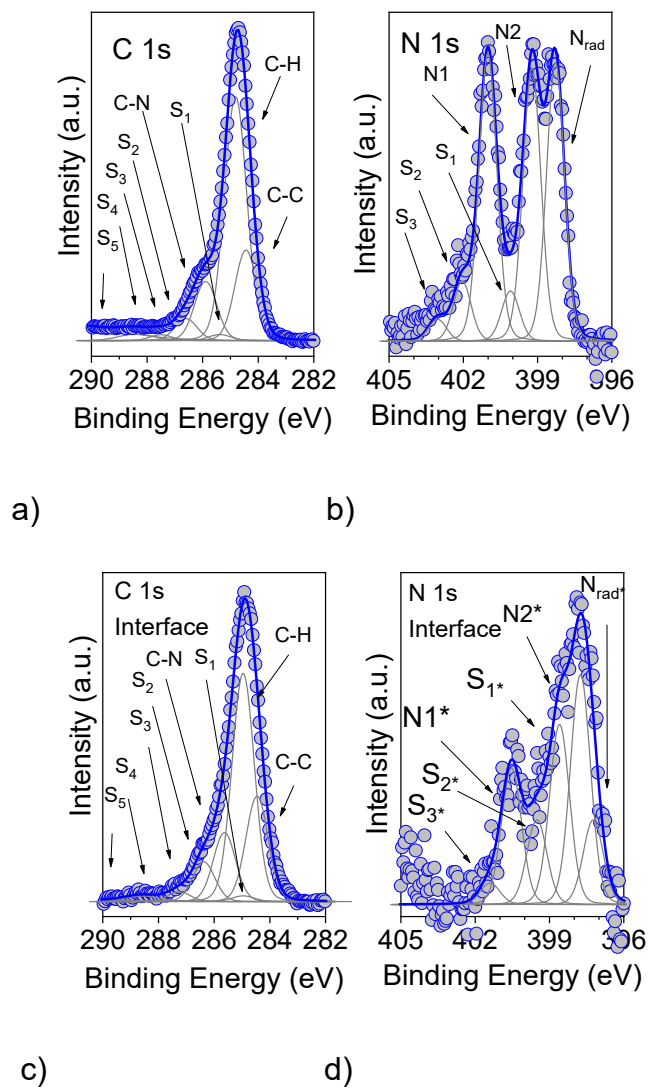

Figure 1. Blatter-pyr films deposited on polycrystalline cobalt. a) C 1s and b) N 1s core-level spectra of a thicker film (6.0 nm) compared to the interfacial layer (0.4 nm) spectra c) and d), together with their fit components, named as shown Blatter-pyr molecular structure. For the fit details see the Supporting Information.

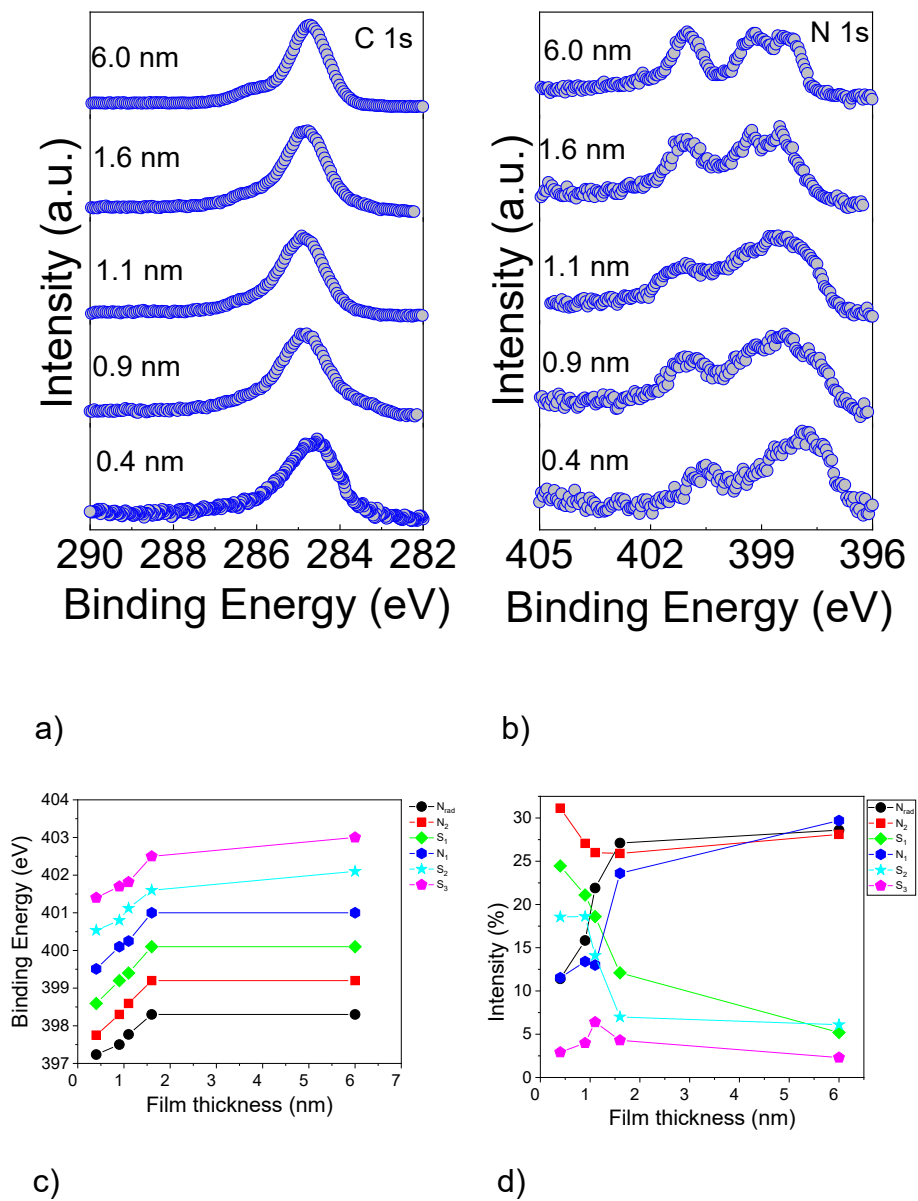

Figure 2. Thickness-dependent a) C 1s and b) N 1s core-level spectra of Blatter-pyr thin films on polycrystalline cobalt. Evolution of c) the energy position and d) the relative intensities of the single components in the N 1s core-level spectra with increasing film thickness, as obtained from the fit procedure applied to the N 1s core-level spectra in b) (details in the Supporting Information).

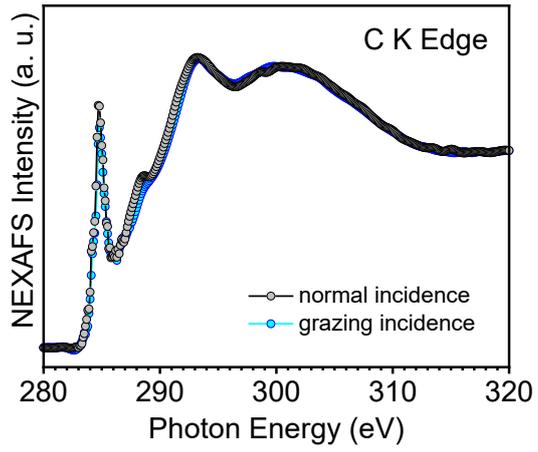 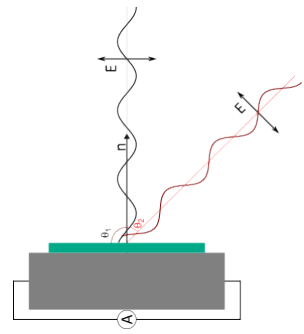

a) b)

Figure 3. a) C K edge of an interfacial Blatter-pyr film deposited on polycrystalline cobalt. b) Geometry of the experiment.

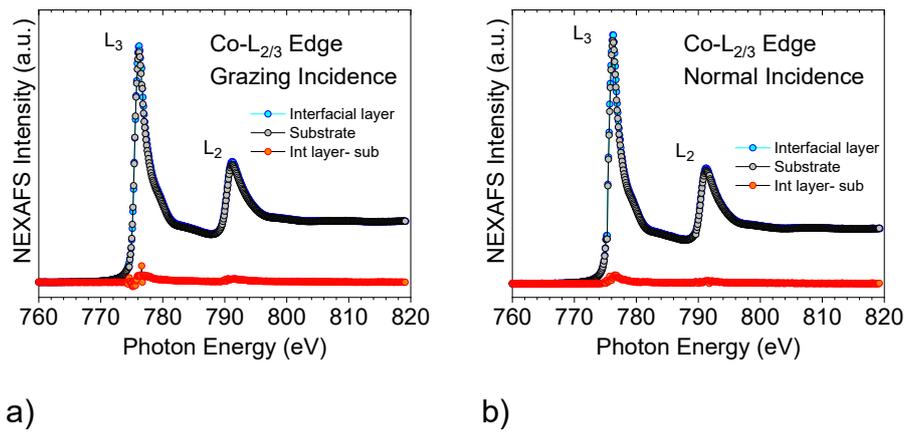

Figure 4. Co L2,3 NEXAFS spectra of the polycrystalline cobalt surface before and after deposition of a layer of Blatter-pyr, as indicated, for a) grazing incidence and b) normal incidence (see also Figure 3).

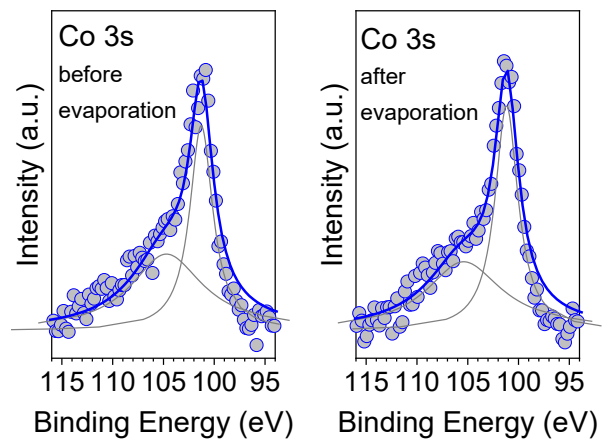

Figure 5. Co 3s core-level spectra a) before and b) after evaporation of the interfacial radical layer, together with their fit.

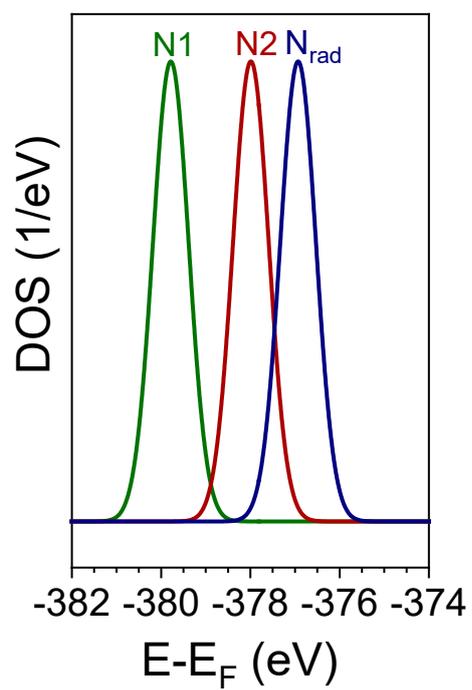

a)

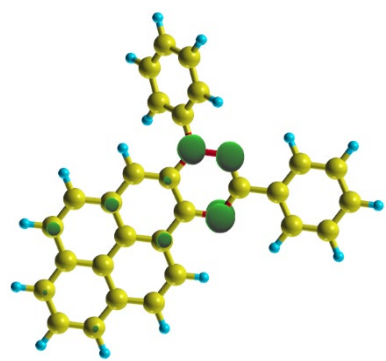

b)

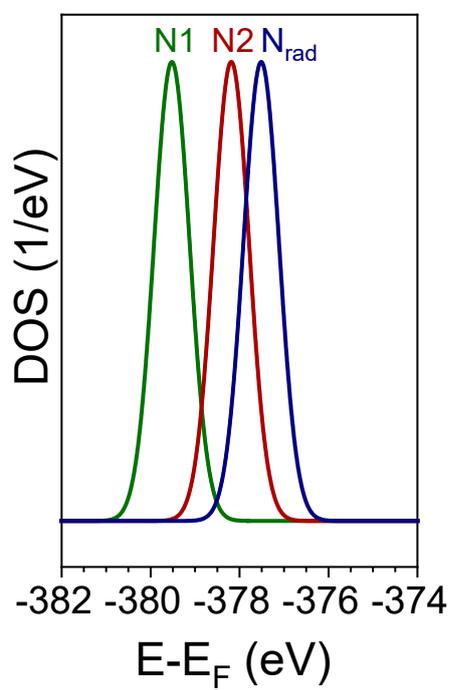

c)

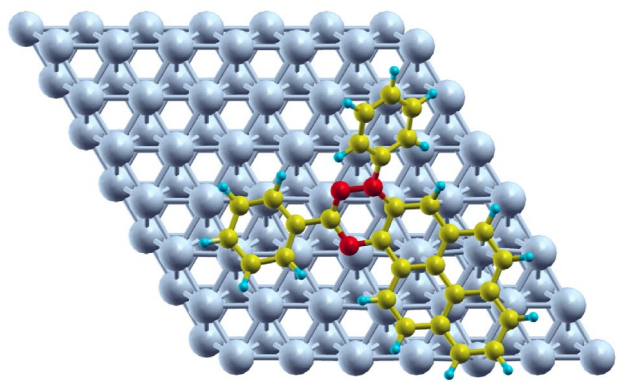

d)

Figure 6. Simulated N 1s XPS spectrum (DOS) for the a) isolated molecule and c) the molecule on the Co surface. The zero-energy reference is set to the Fermi level. b) Spin density isosurface for the isolated molecule. d) Top view of the slab used in the calculations.